\documentclass[11pt,letterpaper]{article}

\usepackage[margin=1in]{geometry}
\usepackage{amsmath,amssymb,amsthm}
\usepackage{graphicx}
\usepackage{algorithm}
\usepackage{algorithmic}
\usepackage{booktabs}
\usepackage{xcolor}
\usepackage{hyperref}
\usepackage{authblk}
\usepackage{listings}
\usepackage{subcaption}

\newtheorem{theorem}{Theorem}[section]

\newtheorem{proposition}[theorem]{Proposition}

\theoremstyle{definition}

\newcommand{\R}{\mathbb{R}}

\title{Exponential divided differences via Chebyshev polynomials}

\author{Itay Hen}
\affil{Information Sciences Institute, University of Southern California, Marina del Rey, CA 90292, USA}
\affil{Department of Physics and Astronomy, University of Southern California, Los Angeles, CA 90089, USA}

\affil{\texttt{itayhen@isi.edu}}

\date{\today}

\begin{document}

\maketitle

\begin{abstract}
\noindent 
Exponential divided differences arise in numerical linear algebra,
matrix-function evaluation, and quantum Monte Carlo simulations, where
they serve as kernel weights for time evolution and observable
estimation. Efficient and numerically stable evaluation of high-order
exponential divided differences for dynamically evolving node sets
remains a significant computational challenge. We present a
Chebyshev-polynomial-based algorithm that addresses this problem by
combining the Chebyshev-Bessel expansion of the exponential function
with a direct recurrence for Chebyshev divided differences. The method
achieves a computational cost of ${\cal O}(qN)$, where $q$ is the
divided-difference order and $N$ is the Chebyshev truncation length. We
show that $N$ scales linearly with the spectral width through the decay
of modified Bessel coefficients, while the dependence on $q$ enters only
through structural polynomial constraints. We further develop an
incremental update scheme for dynamic node sets that enables the
insertion or removal of a single node in ${\cal O}(N)$ time when the
affine mapping interval is held fixed. A full \texttt{C++} reference
implementation of the algorithms described in this work is publicly
available.
\end{abstract}



\vspace{0.3in}

\section{Introduction}

Divided differences are a foundational construct in numerical analysis,
providing a unifying framework for polynomial interpolation, numerical
differentiation, and the computation of derivatives of arbitrary
order~\cite{stoer2002introduction,hildebrand1987introduction}. For a
function $f : \mathbb{R} \to \mathbb{R}$ and distinct nodes
$x_0,\ldots,x_q$, the $q$-th order divided difference is defined
recursively by
\begin{equation}
f[x_0,\ldots,x_q]
=
\frac{f[x_1,\ldots,x_q]-f[x_0,\ldots,x_{q-1}]}{x_q-x_0},
\label{eq:divdiff_def}
\end{equation}
with $f[x_0]=f(x_0)$.
Despite their long history, the efficient and numerically stable
evaluation of divided differences for special functions—particularly at
high order—remains an active area of research~\cite{Zivcovich2019,GuptaAlbashHen2020}. The challenge is most
acute when the nodes exhibit strong clustering, near-degeneracy, complex
values, or a wide spectral extent, conditions under which classical
formulations often fail.

A particularly important case, which arises ubiquitously in numerical linear algebra, matrix-function evaluation, and quantum Monte Carlo (QMC) simulations of
many-body quantum systems~\cite{EzzellBarashHen2025,BabakhaniBarashHen2025,AkaturkHen2024,AlbashWagenbrethHen2017,GuptaHen2020,GuptaAlbashHen2020,BarashBabakhaniHen2024}, is $f(x)=\exp(x)$, for which the exponential divided difference is given by
\begin{equation}
\exp[x_0,x_1,\ldots,x_q] \,.
\label{eq:exp_divdiff}
\end{equation}
In these applications, exponential divided differences appear as
fundamental kernel weights governing imaginary-time propagation,
real-time dynamics, and estimator construction for physical observables.
Crucially, the node sets $\{x_i\}$ are not
static. Instead, they evolve dynamically through incremental insertion
and removal of individual nodes along Monte Carlo trajectories or
sampling paths. In such settings, the computational cost of updating
exponential divided differences—rather than merely evaluating them
once—becomes the dominant algorithmic bottleneck.

The most direct approach to computing divided differences is the
classical divided-difference table~\cite{isaacson1994analysis}, constructed
by iterating Eq.~\eqref{eq:divdiff_def} from the function values
$\exp(x_i)$. While algebraically exact, this representation is
notoriously unstable when nodes are clustered or nearly coincident.
Severe catastrophic cancellation arises from subtracting nearly equal
quantities, obscuring the smooth limiting behavior in which divided
differences converge to derivatives of the exponential. These numerical
pathologies render naive recursion unsuitable for high-order or repeated
evaluations.

A major advance in stable evaluation was achieved by McCurdy, Ng, and
Parlett (MNP)~\cite{McCurdy1984}, who reformulated exponential divided
differences as entries of the matrix exponential of the associated
Opitz matrix~\cite{Opitz1964}. By combining spectral shifting, scaling,
Taylor expansion, and squaring, their algorithm avoids the cancellation
inherent in classical recursion and remains a gold standard for
numerical stability, alongside later refinements~\cite{Zivcovich2019,GuptaAlbashHen2020}. However, the MNP method requires
${\cal O}(q^2 \log c)$ computational work for a node set of size $q$, where
$c \geq [\max(x_i)-\min(x_i)]/2$ denotes (a bound on) the half-width of the inputs. While acceptable for
isolated evaluations, this cost becomes prohibitive in applications
requiring millions of closely related evaluations -- a scenario that arises prominently in quantum
Monte Carlo (PMR) methods. Recomputing a full ${\cal O}(q^2)$ table at each update is therefore
computationally wasteful and quickly dominates the overall simulation
cost.

An important step toward alleviating this inefficiency was taken in Ref.~\cite{GuptaAlbashHen2020}, where a method for the dynamic
addition and removal of inputs based on a
reformulation due to Zivkovich~\cite{Zivcovich2019} was devised, which enabled the efficient updating when nodes are appended to or removed from an
existing set, at a cost on the order of ${\cal O}(qc)$, avoiding full recomputation from scratch, and establishing the practical feasibility of incremental kernel evaluation in Monte
Carlo contexts.

The purpose of this work is to introduce a fundamentally different
computational approach to exponential divided differences based on a
Chebyshev expansion of the exponential function, which offers advantages over existing techniques. Explicitly, we develop and analyze
a numerically stable, implementation-ready algorithm whose cost scales
linearly with the number of nodes for a fixed spectral width. By combining
a Chebyshev expansion with a direct recurrence for Chebyshev divided
differences, the method reduces the cost of a single evaluation to
${\cal O}(q\,N)$, where $q$ is the divided-difference order and $N$ is the
number of Chebyshev terms required for convergence, which we show grows linearly with $c$ and only weakly with $q$ in practically relevant regimes.
In addition, we introduce an incremental update algorithm that enables
efficient modification of exponential divided differences under node insertion and removal, requiring only ${\cal O}(N)={\cal O}(c)$ work per update for a fixed Chebyshev mapping interval.

The remainder of this paper is organized as follows.
Section~\ref{sec:chebyshev} presents the Chebyshev polynomial–based algorithm, including the Chebyshev–Bessel expansion, the
associated divided-difference recurrence, and practical implementation details.
Section~\ref{sec:runtime_analysis} analyzes the computational complexity of the method.
Numerical validation of correctness, stability, and the predicted scaling behavior is presented
in Sec.~\ref{sec:numerical}.
Section~\ref{sec:incremental} introduces an incremental update scheme for dynamically evolving
node sets, demonstrating how individual node insertions and removals can be handled efficiently
without full recomputation and in Sec.~\ref{sec:beta_bessel} we show that scaled exponential divided differences
\(\exp(-\beta[x_0,\ldots,x_q])\) can be evaluated within the same Chebyshev--Bessel framework
via a simple rescaling of the nodes.
Finally, Sec.~\ref{sec:conclusions} summarizes the main results and discusses implications and
possible extensions of the approach.

\section{Chebyshev polynomial method}
\label{sec:chebyshev}

We next derive a Chebyshev polynomial–based algorithm
for the stable and efficient evaluation of exponential divided differences.
The method maps the input nodes to the canonical Chebyshev interval
$[-1,1]$ and combines the Chebyshev--Bessel expansion of the exponential
function with a direct recurrence for Chebyshev divided differences.
This construction avoids the catastrophic cancellation inherent in classical
divided-difference tables and yields a computational cost that scales linearly
with the divided-difference order for fixed spectral width. We begin by introducing the necessary mathematical machinery.

\subsection{Divided differences}

For a function $f : \R \to \R$ and distinct nodes
$x_0,\ldots,x_q \in \R$, the divided difference $f[x_0,\ldots,x_q]$
admits several equivalent characterizations.

\begin{proposition}[Basic properties of divided differences]
\label{prop:divdiff_properties}
The divided difference satisfies:
\begin{enumerate}
\item \textbf{Linearity}:
$(\alpha f + \beta g)[x_0,\ldots,x_q]
= \alpha f[x_0,\ldots,x_q] + \beta g[x_0,\ldots,x_q]$.
\item \textbf{Symmetry}:
$f[x_0,\ldots,x_q]$ is invariant under permutations of the nodes.
\item \textbf{Leibniz rule}:
\[
(fg)[x_0,\ldots,x_q]
=
\sum_{k=0}^q
f[x_0,\ldots,x_k]\,
g[x_k,\ldots,x_q].
\]
\item \textbf{Smoothness}:
If $f \in C^q$, then
$f[x_0,\ldots,x_q] = f^{(q)}(\xi)/q!$
for some
$\xi \in \mathrm{conv}(x_0,\ldots,x_q)$.
\end{enumerate}
\end{proposition}

Another key technical ingredient of our technique is the  behavior
of divided differences under affine changes of variables.
Let $x = d + c y$ with $c \neq 0$. Then
\begin{equation}
f(d+cy)[x_0,\ldots,x_q]
=
\frac{1}{c^q}\,
g[y_0,\ldots,y_q],
\label{eq:jacobian}
\end{equation}
where $g(y)=f(d+cy)$ and $y_i=(x_i-d)/c$.

\begin{proof}
The claim follows by induction on $q$.
For $q=1$,
\[
f[x_0,x_1]
=
\frac{f(x_1)-f(x_0)}{x_1-x_0}
=
\frac{g(y_1)-g(y_0)}{c(y_1-y_0)}
=
\frac{1}{c}\, g[y_0,y_1].
\]
Assuming the result holds for order $q-1$, the recursive definition of
divided differences yields
\[
f[x_0,\ldots,x_q]
=
\frac{c^{-(q-1)}\bigl(g[y_1,\ldots,y_q]-g[y_0,\ldots,y_{q-1}]\bigr)}
{c(y_q-y_0)}
=
\frac{1}{c^q}\, g[y_0,\ldots,y_q].
\]
\end{proof}

\subsection{Chebyshev--Bessel expansion of the exponential}

\begin{figure}[htp]
  \centering
    \includegraphics[width=0.88\linewidth]{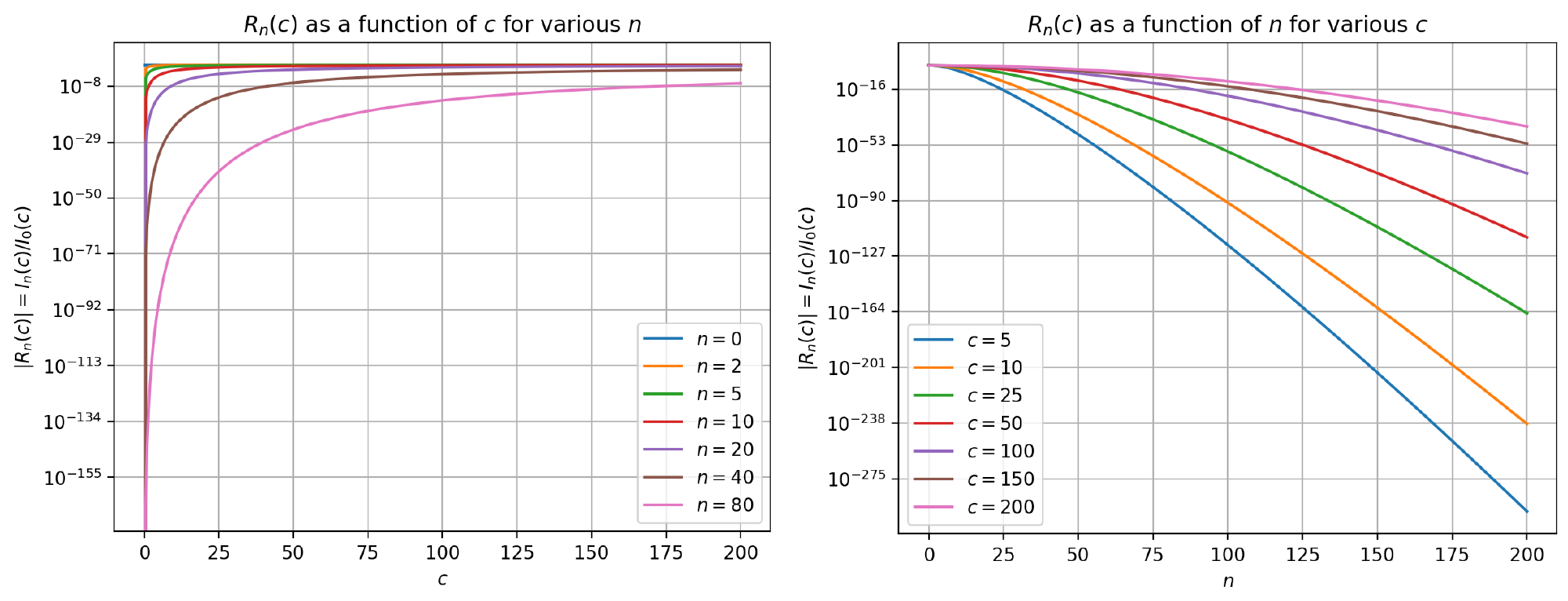}
  \caption{
  Modified-Bessel ratio coefficients \(R_n(c)=I_n(c)/I_0(c)\) governing the Chebyshev--Bessel expansion of \(e^{cy}\) on \(y\in[-1,1]\). Left: the dependence on \(c\) for representative orders \(n\). Right: the decay in \(n\) for representative \(c\) values, illustrating the concentration of weight around \(n=\Theta(c)\) that drives the truncation length \(N=\Theta(c)\) at fixed tolerance.  }
  \label{fig:Rn_panels}
\end{figure}

The exponential function admits a Chebyshev expansion on $[-1,1]$.
For $y\in[-1,1]$ and $c\in\R$,
\begin{equation}
\exp(cy)
=
I_0(c)
+
2\sum_{n=1}^\infty
I_n(c)\, T_n(y),
\label{eq:cheb_bessel}
\end{equation}
where $T_n$ is the Chebyshev polynomial of the first kind and
$I_n$ is the modified Bessel function of the first kind.
Factoring out the dominant exponential growth yields
\begin{equation}
\exp(cy)
=
I_0(c)
\left(
1
+
2\sum_{n=1}^\infty
R_n(c)\, T_n(y)
\right),
\qquad 
\textrm{where}\quad 
R_n(c)=\frac{I_n(c)}{I_0(c)}.
\label{eq:cheb_besselRat}
\end{equation}
As this will become important shortly, we note here that the ratios $R_n(c)$ satisfy $0\le R_n(c)\le1$ and decay monotonically in $n$ for fixed $c$ (see Fig.~\ref{fig:Rn_panels}). This decay rate will control the truncation length of the
Chebyshev series and, consequently, the computational cost of the algorithm. 

The Chebyshev polynomials obey the three-term recurrence
\begin{equation}
T_{n+1}(y)
=
2y\,T_n(y)
-
T_{n-1}(y),
\qquad
T_0(y)=1,\quad T_1(y)=y.
\label{eq:cheb_recurrence}
\end{equation}
One central algorithmic insight we utilize is that Chebyshev divided differences
can be computed directly from the recurrence~\eqref{eq:cheb_recurrence},
without evaluating $T_n(y_i)$ at the nodes.

\begin{theorem}[Chebyshev divided-difference recurrence]
\label{thm:cheb_divdiff_recurrence}
For any nodes $y_0,\ldots,y_k\in[-1,1]$ with $k\geq 1$,
\begin{equation}
T_{n+1}[y_0,\ldots,y_k]
=
2\Bigl(
y_k\,T_n[y_0,\ldots,y_k]
+
T_n[y_0,\ldots,y_{k-1}]
\Bigr)
-
T_{n-1}[y_0,\ldots,y_k].
\label{eq:cheb_divdiff_rec}
\end{equation}
\end{theorem}

\begin{proof}
Applying the divided-difference operator to
Eq.~\eqref{eq:cheb_recurrence} gives
\[
T_{n+1}[y_0,\ldots,y_k]
=
2(y\,T_n)[y_0,\ldots,y_k]
-
T_{n-1}[y_0,\ldots,y_k].
\]
Using the Leibniz rule and the fact that higher divided differences of
linear functions vanish,
\[
(y\,T_n)[y_0,\ldots,y_k]
=
y_k\,T_n[y_0,\ldots,y_k]
+
T_n[y_0,\ldots,y_{k-1}],
\]
which completes the proof.
\end{proof}

This recurrence avoids the evaluation of complete homogeneous symmetric
polynomials and eliminates the overflow and cancellation issues that
plague other approaches. Moreover, it requires only the divided
differences of $T_0$ and $T_1$. Since $T_0$ is constant and $T_1$ is linear, we have
\begin{align}
T_0[y_0,\ldots,y_k]
&=
\begin{cases}
1, & k=0,\\
0, & k>0,
\end{cases}
\\[4pt]
T_1[y_0,\ldots,y_k]
&=
\begin{cases}
y_0, & k=0,\\
1, & k=1,\\
0, & k>1.
\end{cases}
\end{align}
These values follow directly from the general theory of divided differences.

\subsection{The algorithm}
\label{subsec:algorithm}

Based on the preceding observations, we now present an algorithm for
computing the exponential divided difference $\exp[x_0,\ldots,x_q]$.
Let $[a,b]$ denote an interval containing all nodes, and define
\[
c=\frac{b-a}{2},
\qquad
d=\frac{b+a}{2},
\quad
\text{together with the affine map}
\quad
y_i=\frac{x_i-d}{c},
\]
so that $y_i\in[-1,1]$ for all $i$.
With this normalization, the target quantity can be written as
\begin{equation}
\exp[x_0,\ldots,x_q]
=
\frac{e^d I_0(c)}{c^q}
\left(
D_q^{(0)}
+
2\sum_{n=1}^N
R_n(c)\,D_q^{(n)}
\right),
\label{eq:exp_dd_cheb_normalized}
\end{equation}
where $D_q^{(n)} = T_n[y_0,\ldots,y_q]$ and $N$ denotes the number of terms
retained in the expansion. Since $b-a$ bounds the range of the inputs, we
refer to $c$ as the corresponding half-width.

We next (pre)compute ratio coefficients $R_n(c)$ with $n=0 \ldots N_{\max}$ where $N_{\max}$ bounds $N$ from above (we discuss the scaling of $N$ in detail in the next section). In most applications where exponential divided differences arise, the function must be evaluated repeatedly for different node sets whose values lie within a known interval. In such settings, the interval half-width $c$ is fixed, and all quantities that depend only on $c$ can therefore be computed once and reused across subsequent evaluations.
The precomputation cost is $\mathcal{O}(N_{\max})$ and is amortized over all subsequent evaluations. This computation can be achieved by Miller’s backward recurrence~\cite{DLMF10.41}, which has the important property of producing an entire sequence of modified Bessel ratios in a single computation. Starting from a sufficiently large order and propagating the three-term recurrence downward, the algorithm generates values proportional to $I_n(c)$ for all $n$. After normalization by the computed $I_0(c)$, this yields the full set of ratios simultaneously for all $n$, with no additional recurrence or special-function evaluations. The computational cost scales linearly with the number of terms and is therefore independent of the number of ratios ultimately retained, making Miller’s algorithm particularly efficient when a contiguous block of ratios up to a prescribed cutoff is required.

The Chebyshev divided differences are computed using
Eq.~\eqref{eq:cheb_divdiff_rec} with rolling storage.
Only three Chebyshev layers are required at any time. 
The complete procedure is summarized in
Algorithm~\ref{alg:chebyshev_fixed_c}.
The series is truncated adaptively at some $N$ so
that the remaining tail is below a prescribed relative tolerance $\varepsilon$.

\begin{algorithm}
\caption{Chebyshev method for exponential divided differences (fixed $c$)}
\label{alg:chebyshev_fixed_c}
\begin{algorithmic}[1]
\REQUIRE Nodes $x_0,\ldots,x_q$, precomputed ratios $R_n(c)$, tolerance $\varepsilon$
\ENSURE Approximation to $\exp[x_0,\ldots,x_q]$
\STATE Map nodes: $y_i \gets (x_i-d)/c$ for $i=0,\ldots,q$
\STATE Initialize $D_k^{(0)} \gets \delta_{k,0}$ for $k=0,\ldots,q$
\STATE Initialize $D_0^{(1)} \gets y_0$, \; $D_1^{(1)} \gets 1$, \; $D_k^{(1)} \gets 0$ for $k \geq 2$
\STATE $S \gets D^{(0)}_q + 2R_1(c)\,D^{(1)}_q$
\FOR{$n = 2$ to $N_{\max}$}
    \FOR{$k = 0$ to $q$}
        \STATE $D_k^{(n)} \gets 2\bigl(y_k D_k^{(n-1)} + D_{k-1}^{(n-1)}\bigr) - D_k^{(n-2)}$ \hfill \COMMENT{with $D_{-1}^{(n-1)} \equiv 0$}
    \ENDFOR
    \STATE $\mathrm{term} \gets 2R_n(c)\, D^{(n)}_q$
    \STATE $S \gets S + \mathrm{term}$
    \IF{$|\mathrm{term}| < \varepsilon |S|$}
        \STATE \textbf{break}
    \ENDIF
    \STATE Shift layers: $D^{(n-2)}\gets D^{(n-1)}$, $D^{(n-1)}\gets D^{(n)}$
\ENDFOR
\RETURN $e^{d} I_0(c)\, S / c^q$
\end{algorithmic}
\end{algorithm}

\section{Runtime analysis}
\label{sec:runtime_analysis}

We now analyze the computational cost of evaluating exponential divided
differences using the Chebyshev--Bessel expansion and the Chebyshev
divided-difference recurrence introduced in
Section~\ref{sec:chebyshev}.
The total runtime admits a clean decomposition into two conceptually
distinct components:
(i)~the cost of advancing the Chebyshev recurrence by one polynomial
order, and
(ii)~the number of Chebyshev terms \(N(q,c;\varepsilon)\) required to
achieve a prescribed accuracy.
The overall cost is essentially the product of these two contributions.

\subsection{Per-term cost: ${\cal O}(q)$ work per Chebyshev order}
\label{sec:pertermcost}

The core recurrence advances Chebyshev divided differences in the
Chebyshev index \(n\).
For each \(n\), the algorithm maintains a single layer
\begin{equation}
\mathbf D^{(n)}
=
\bigl(D_0^{(n)},D_1^{(n)},\ldots,D_q^{(n)}\bigr),
\qquad
D_k^{(n)}:=T_n[y_0,\ldots,y_k],
\end{equation}
and updates \(\mathbf D^{(n)}\) from \(\mathbf D^{(n-1)}\) and
\(\mathbf D^{(n-2)}\) via a single forward sweep over \(k=0,\ldots,q\).
Each update requires a constant number of arithmetic operations per
entry.
Consequently, advancing the Chebyshev order \(n\mapsto n{+}1\) incurs
\({\cal O}(q)\) work.
Evaluating the corresponding contribution to the Chebyshev--Bessel
series further requires only a single multiplication by the scalar
coefficient \(R_n(c)\) and an accumulation into the running sum, adding
\({\cal O}(1)\) additional work.
If \(N\) Chebyshev orders are required to reach convergence, the total
runtime of a single evaluation therefore satisfies
\begin{equation}
T_{\rm eval}(q,c;\varepsilon)
=
{\cal O}(q)\,N(q,c;\varepsilon)
+
{\cal O}(q),
\label{eq:runtime_basic}
\end{equation}
where the final \({\cal O}(q)\) term accounts for node mapping,
initialization, and the final scaling factor \(e^{d}/c^{q}\).
In all practically relevant regimes, the dominant cost is
\({\cal O}(qN)\).

As already mentioned, the ratio coefficients \(R_n(c)\) may either be precomputed once for a
fixed interval half-width \(c\), at cost \({\cal O}(N)\) amortized across
many evaluations, or generated on the fly via a stable recurrence at
\({\cal O}(1)\) cost per \(n\).
Either choice leaves the asymptotic \({\cal O}(qN)\) scaling unchanged.

\subsection{Truncation bounds for the Chebyshev--Bessel expansion of exponential divided differences}
\label{sec:truncation_bound}

Having established that each Chebyshev order costs ${\cal O}(q)$ work, we now determine the number of terms $N$ required for convergence.
We begin from the Chebyshev--Bessel representation
\begin{equation}
\exp[x_0,\ldots,x_q]
=
\frac{e^{d} I_0(c)}{c^{q}}
\left(
D_q^{(0)} + 2\sum_{n=1}^{\infty} R_n(c)\, D_q^{(n)}
\right),
\label{eq:cheb_bessel_dd_expansion_infinite}
\end{equation}
where
\begin{equation}
D_q^{(n)} = T_n[y_0,\ldots,y_q],
\qquad y_j\in[-1,1],
\label{eq:Dqn_def}
\end{equation}
and \(R_n(c)=I_n(c)/I_0(c)\) are normalized modified Bessel coefficients.
Truncating the series at index \(N\) yields
\begin{equation}
\exp_N[x_0,\ldots,x_q]
=
\frac{e^{d} I_0(c)}{c^{q}}
\left(
D_q^{(0)} + 2\sum_{n=1}^{N} R_n(c)\, D_q^{(n)}
\right).
\label{eq:cheb_bessel_dd_expansion_truncated}
\end{equation}
The corresponding truncation error is
\begin{equation}
\mathcal{E}_N
=
\left|\exp[x_0,\ldots,x_q]-\exp_N[x_0,\ldots,x_q]\right|
=
\frac{2e^{d} I_0(c)}{c^{q}}
\left|
\sum_{n>N} R_n(c)\, D_q^{(n)}
\right|.
\label{eq:EN_start}
\end{equation}
Applying the triangle inequality gives the basic absolute error bound
\begin{equation}
\mathcal{E}_N
\le
\frac{2e^{d} I_0(c)}{c^{q}}
\sum_{n>N} |R_n(c)|\, |D_q^{(n)}|.
\label{eq:EN_triangle}
\end{equation}

\subsubsection{Uniform bounds on Chebyshev divided differences}

To control the truncation error, we first bound the Chebyshev divided
differences \(D_q^{(n)}=T_n[y_0,\ldots,y_q]\) uniformly over all node sets
\(y_0,\ldots,y_q\in[-1,1]\).
Polynomial degree considerations impose the exact structural constraint
\begin{equation}
T_n[y_0,\ldots,y_q]=0,
\qquad n<q,
\label{eq:DD_zero}
\end{equation}
which holds independently of node placement and tolerance.
As a result, the truncation index must satisfy the hard lower bound
\(N\ge q\), and only terms with \(n\ge q\) contribute.

Assuming distinct nodes (the general case follows by continuity), the
mean-value form of divided differences implies that for any
\(f\in C^q([-1,1])\),
\begin{equation}
f[y_0,\ldots,y_q]
=
\frac{f^{(q)}(\xi)}{q!},
\qquad \xi\in(-1,1),
\label{eq:DD_mvt}
\end{equation}
and hence
\begin{equation}
|f[y_0,\ldots,y_q]|
\le
\frac{1}{q!}\max_{x\in[-1,1]} |f^{(q)}(x)|.
\label{eq:DD_mvt_bound}
\end{equation}
Applying this bound to \(f(x)=T_n(x)\) yields
\begin{equation}
|T_n[y_0,\ldots,y_q]|
\le
\frac{1}{q!}\max_{x\in[-1,1]} |T_n^{(q)}(x)|.
\label{eq:TnDD_to_derivative}
\end{equation}

The \(q\)th derivative of the Chebyshev polynomial admits the exact
representation
\begin{equation}
\frac{d^q}{dx^q}T_n(x)
=
2^{q-1}\,n\,(q-1)!\, C^{(q)}_{n-q}(x),
\qquad n\ge q,
\label{eq:Tnq_derivative_gegenbauer}
\end{equation}
where \(C^{(\lambda)}_m(x)\) denotes a Gegenbauer polynomial.
For \(\lambda>0\), the sharp supremum norm identity
\begin{equation}
\max_{x\in[-1,1]} |C^{(\lambda)}_m(x)|
=
C^{(\lambda)}_m(1)
=
\frac{(2\lambda)_m}{m!}
\label{eq:gegenbauer_sup}
\end{equation}
holds.
Combining these results gives the uniform bound
\begin{align}
|D_q^{(n)}|
&=
|T_n[y_0,\ldots,y_q]|
\le
\frac{1}{q!}\,2^{q-1}\,n\,(q-1)!\,
\frac{(2q)_{n-q}}{(n-q)!}
\nonumber\\
&=
2^{q-1}\,\frac{n}{q}\,\frac{(2q)_{n-q}}{(n-q)!},
\qquad n\ge q.
\label{eq:Dqn_uniform_bound_exact}
\end{align}
For fixed \(q\) and large \(n\),
\[
\frac{(2q)_{n-q}}{(n-q)!}
=
\frac{\Gamma(n+q)}{\Gamma(2q)\Gamma(n-q+1)}
\sim
\frac{n^{2q-1}}{\Gamma(2q)},
\]
and therefore
\begin{equation}
|D_q^{(n)}|
\le
\frac{2^{q-1}}{q\,\Gamma(2q)}\, n^{2q}\,(1+{\cal O}(1)).
\label{eq:Dqn_asymptotic}
\end{equation}

\subsubsection{Tail bounds and relative truncation criteria}

Substituting the uniform bound
\eqref{eq:Dqn_uniform_bound_exact} into
Eq.~\eqref{eq:EN_triangle} yields
\begin{equation}
\mathcal{E}_N
\le
\frac{2e^{d} I_0(c)}{c^{q}}
\sum_{n>\max\{N,q-1\}}
|R_n(c)|\,
2^{q-1}\,\frac{n}{q}\,\frac{(2q)_{n-q}}{(n-q)!}.
\label{eq:EN_bound_exact}
\end{equation}
Using the large-\(n\) envelope \eqref{eq:Dqn_asymptotic}, we obtain the
simplified scaling bound
\begin{equation}
\mathcal{E}_N
\;\lesssim\;
\frac{2e^{d} I_0(c)}{c^{q}}
\sum_{n>N}
\frac{I_n(c)}{I_0(c)}\frac{n^{2q}}{q!},
\label{eq:EN_bound_simplified}
\end{equation}
where the implicit constant depends only on \(q\) and does not affect the
asymptotic dependence on \(n\), \(c\), or \(\varepsilon\).

To convert this absolute bound into a relative truncation criterion, we
introduce the nonnegative coefficients
\begin{equation}
a_n := \frac{I_n(c)}{I_0(c)}\frac{n^{2q}}{q!},
\qquad
S := \sum_{n\ge 0} a_n,
\qquad
S_N := \sum_{n=0}^{N} a_n.
\label{eq:def_an_S}
\end{equation}
Up to the common prefactor \(2e^{d} I_0(c)/c^{q}\), both the full expansion
and its truncation are governed by the same coefficient sequence
\(\{a_n\}\).
A sufficient condition for relative accuracy is therefore
\begin{equation}
\frac{\mathcal{E}_N}{\exp[x_0,\ldots,x_q]}
\;\lesssim\;
\frac{S-S_N}{S}
\;\le\;
\varepsilon.
\label{eq:relative_tail_condition}
\end{equation}
The behavior of the relative tail \((S-S_N)/S\) depends on which portion
of the modified-Bessel profile dominates the sum.
When \(c\gg q\), the profile is sharply concentrated near
\(n\approx c\) with width \({\cal O}(\sqrt{c})\), and the polynomial factor
\(n^{2q}\) varies slowly across this window.
Approximating the Bessel profile by a Gaussian yields
\begin{equation}
N
\approx
c + \sqrt{2c\log\!\frac{1}{\varepsilon}}.
\label{eq:N_rel_cggq}
\end{equation}
In contrast, when \(q\gg c\), the polynomial weight shifts the dominant
contribution into the factorial decay regime of the Bessel function.
A saddle-point analysis yields a peak at
\begin{equation}
n_\star \approx \frac{2q}{W(4q/c)},
\label{eq:nstar_lambertW}
\end{equation}
and placing the truncation index beyond this peak gives
\begin{equation}
N
\approx
\frac{2q}{W(4q/c)}
+
\sqrt{2q\log\!\frac{1}{\varepsilon}}.
\label{eq:N_rel_qggc}
\end{equation}
These estimates summarize the relative truncation behavior in the
width-controlled and order-controlled regimes.
In practice, they provide reliable guidance for choosing \(N\), while
adaptive ratio-based tail tests offer a robust alternative when a priori regime classification is undesirable.
\section{Numerical verification}
\label{sec:numerical}

The purpose of this section is to numerically verify the correctness of the algorithm derived in Sec,~\ref{sec:chebyshev} and the bounds established in Sec.~\ref{sec:runtime_analysis}. 
%
All numerical data presented in this section were generated using a single, deterministic implementation of the Chebyshev divided-difference algorithm described in Sec.~\ref{subsec:algorithm}. For each data point, the relevant control parameters (divided-difference order \(q\), the interval \([a,b]\), and convergence tolerance $\varepsilon$) were fixed explicitly, while random node sets \(\{x_0,\ldots,x_q\}\subset[a,b]\) were drawn from a uniform distribution. Reported values were obtained by calculating the median across multiple (generally $100$) independent realizations. 

\subsection{Comprehensive Verification of the Chebyshev Method}
\label{sec:numerical_correctness}

To ensure the reliability and accuracy of our technique, we conducted an extensive verification campaign that compared the algorithm against the established MNP algorithm~\cite{McCurdy1984}. This validation strategy was designed to test the implementation across a wide range of parameter regimes and edge cases, combining component-level validation with systematic head-to-head comparisons and stress testing at extreme parameters.
We tested multiple values of the interval half-width $c$, ranging from 0.1 to 500, and crossed them with multiple values of the divided difference order $q$, from 1 to 500, using 100 randomly generated node configurations for each $(q,c)$ combination. This yielded tens of thousands of individual divided difference evaluations, allowing us to map out agreement patterns across four orders of magnitude in interval size and two orders of magnitude in derivative order. 

We found that the algorithm not only matches NNP's accuracy where both succeed but extends functionality into parameter regimes where MNP fails  (this is further discussed in Sec.~\ref{sec:mccurdy_comparison}), demonstrating a 100\% success rate across all tested configurations.


\subsection{Computational efficiency and convergence characteristics}
\label{sec:efficiency}

Section~\ref{sec:pertermcost} establishes that the computational cost of a single evaluation using the Chebyshev method scales linearly with the number of retained Chebyshev terms \(N\), with a per-term cost proportional to the divided-difference order \(q\). 
A second central result of the previous section is that the truncation length \(N\)
is governed by the competition between the modified Bessel envelope \(R_n(c)\) [set by the
interval half-width \(c=(b-a)/2\)] and the growth of the Chebyshev divided differences with \(n\),
together with the structural constraint \(D_q^{(n)}\equiv 0\) for \(n<q\), which enforces the hard
lower bound \(N\ge q\).
In the width-dominated regime \(c\gg q\), the decay of \(R_n(c)\) controls truncation, and one
finds \(N=\Theta(c)\) at fixed tolerance, with only weak residual dependence on \(q\) beyond the
constraint \(N\ge q\). In contrast, when \(q\gg c\), the order dependence becomes explicit, and the truncation scale
grows with \(q\).

To verify these predictions numerically, we measure the minimal truncation length \(N\) required
to satisfy the convergence criterion as a function of both the interval half-width \(c\) and the
divided-difference order \(q\).
For each data point, we generate random node sets
\(\{x_0,\ldots,x_q\}\subset[a,b]\) and execute the full Chebyshev recurrence until the convergence
condition is met, at which point the corresponding truncation index \(N\) is recorded.
This procedure is repeated across a range of divided-difference orders spanning both small-\(q\)
and large-\(q\) regimes, as well as across multiple values of the interval half-width \(c\).
Reported values of \(N\) are taken as the median over $100$ independent realizations to suppress
statistical fluctuations and ensure robustness with respect to node placement.
The study was done over a two-dimensional parameter grid comprising
multiple values of the half-width \(c\) between 0.05 and 100 (spanning nearly four orders of magnitude)
and multiple values of the divided-difference order \(q\) between 1 and 100,  recording the number of Chebyshev terms required to satisfy the purely relative
convergence criterion \(|\text{term}| < \varepsilon |S|\) with \(\varepsilon = 10^{-14}\).
In total, this yielded tens of thousands of individual convergence measurements, enabling a detailed mapping
of the convergence landscape and the extraction of empirical scaling behavior.

\begin{figure}[htbp]
    \centering
    \includegraphics[width=0.84\textwidth]{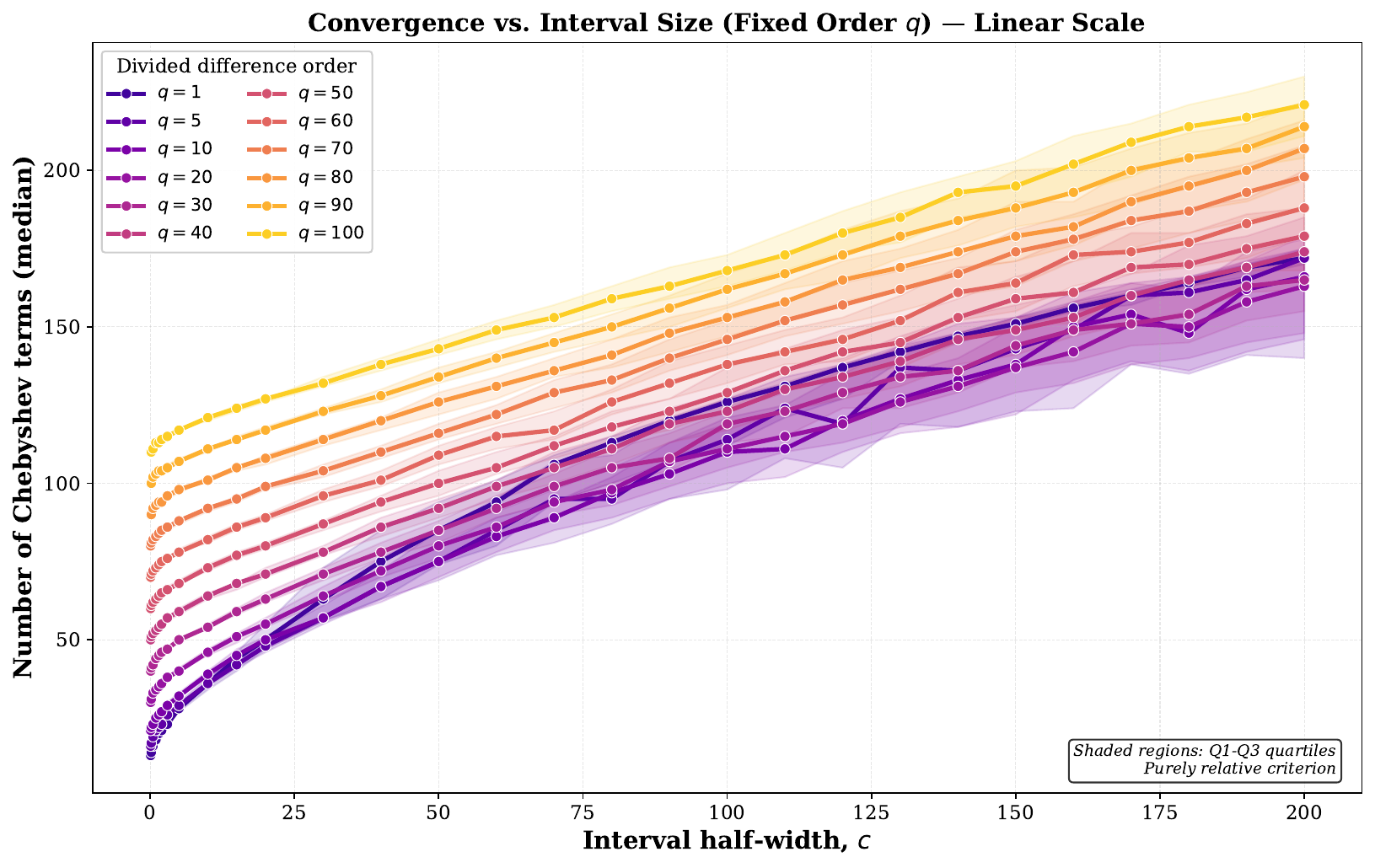}
    \caption{%
        Median number of Chebyshev terms required for convergence as a function of interval half-width $c$ for various fixed divided difference orders $q$.
        Lines show the median (Q2) over 100 random trials per $(c,q)$ combination, with shaded regions indicating the interquartile range (Q1-Q3).
        The linear scale presentation reveals that convergence requirements grow sub-linearly with interval size, with high-order curves ($q \geq 50$) exhibiting particularly weak $c$-dependence.
        For small intervals ($c \lesssim 1$), all curves converge rapidly with $N_{\mathrm{terms}} \approx q + 10-15$ nearly independent of $c$.
        At large intervals ($c \gtrsim 50$), the curves fan out but maintain modest growth, with the highest tested order ($q=100$) requiring only 170 terms even at $c=200$.
        The narrow quartile ranges demonstrate excellent consistency across random node configurations.
    }
    \label{fig:convergence_vs_c_fixed_q}
\end{figure}

Figure~\ref{fig:convergence_vs_c_fixed_q} presents the median number of required Chebyshev terms as a function of interval half-width $c$ for various fixed divided difference orders $q$. The linear scale presentation reveals approximately linear dependence on $c$ with a subleading $\sqrt{c}$ correction (consistent with Sec. 3), which appears as mild concavity on the plotted range, with the curves fanning out at large intervals but remaining remarkably flat at small $c$. For small intervals where $c \lesssim 1$, all curves collapse toward a simple constant offset behavior with $N_{\text{terms}} \approx q + 10-15$, nearly independent of $c$ (recall that $D_q^{(n)}=0$ identically for $n<q$ and Bessel ratios decay rapidly for small $c$.
This weak $c$-dependence becomes even more pronounced at high orders: the $q=100$ curve increases by only 10\% (from 110 to 121 terms) as $c$ varies from 0.1 to 10, a hundred-fold change in interval size. The physical origin of this behavior lies in the triangular structure of the Chebyshev recurrence, which ensures that the first $q$ terms contribute identically zero, leaving convergence dominated by the rapid decay of Bessel ratios $R_n(c)$ for small $c$. As interval size increases, the curves exhibit modest growth but remain well-behaved even at the extreme tested configuration of $c=200, q=100$, which requires only 170 terms. The complementary behavior—weak $q$-dependence at large $c$—is demonstrated in the regime analysis of Figure~\ref{fig:regime_analysis}(a), where curves at fixed large intervals ($c=50, 100, 200$) show that a thirty-fold increase in divided difference order produces only mild increases in required terms, confirming that convergence in the large-interval regime is dominated by the slow decay of Bessel ratios rather than the derivative order.

\begin{figure}[htbp]
    \centering
    \includegraphics[width=0.84\textwidth]{panel2.pdf}
    \caption{%
        Convergence regime analysis demonstrating weak parameter dependencies in extreme limits.
        (a) Large-interval regime ($c \gg q$): For fixed large intervals ($c=50, 100, 200$), convergence exhibits mild $q$-dependence.
        A 30-fold increase in divided difference order ($q: 1 \to 30$) produces only 9--16\% increases in required terms.
        The flat curves reflect a saturation effect where slow Bessel ratio decay at large $c$ dominates the convergence requirement, rendering the method nearly insensitive to derivative order in this regime.
        (b) High-order regime ($q \gg c$): For fixed high orders ($q=50, 70, 100$), convergence exhibits mild $c$-dependence.
        A 100-fold increase in interval size ($c: 0.1 \to 10$) produces only 10--22\% increases in required terms, with weaker sensitivity at higher $q$.
        The near-horizontal curves confirm that when $q$ is large, the triangular structure of the Chebyshev recurrence (which ensures $D_n[q]=0$ for $n<q$) dominates convergence behavior independent of interval size.
    }
    \label{fig:regime_analysis}
\end{figure}

\begin{figure}[htbp]
    \centering
    \includegraphics[width=0.84\textwidth]{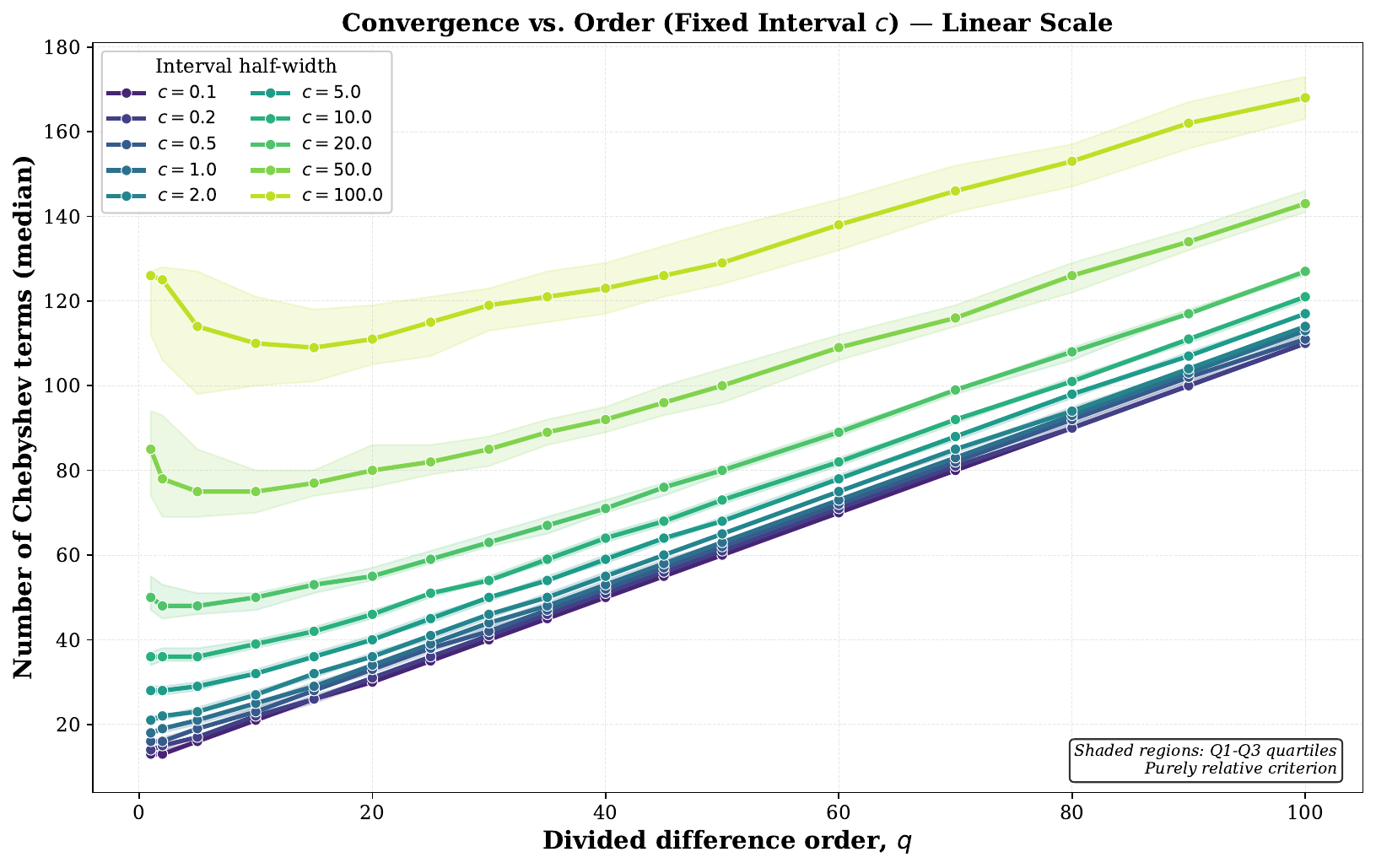}
    \caption{%
        Median number of Chebyshev terms required for convergence as a function of divided-difference order \(q\) for various fixed interval half-widths \(c\).
        Lines show the median (Q2) over 100 random trials per \((c,q)\) combination, with shaded regions indicating the interquartile range (Q1--Q3).
        The linear-scale presentation reveals an approximately linear relationship.
        For larger intervals (\(c \ge 50\)), the curves exhibit approximately linear growth with a subleading square-root correction, reflecting weaker \(q\)-dependence in the regime where interval size dominates convergence behavior.
        The small quartile spread across all parameter combinations confirms that convergence is determined primarily by \((c,q)\) rather than specific node positions.%
    }
    \label{fig:convergence_vs_q_fixed_c}
\end{figure}

Figure~\ref{fig:convergence_vs_q_fixed_c} presents the complementary view, showing median convergence versus divided difference order $q$ for fixed interval half-widths $c$. The linear scale presentation clearly reveals the fundamental linear relationship governing convergence at small intervals. The narrow quartile ranges, barely visible as thin shaded bands around each curve, confirm that this linear relationship holds consistently across random node configurations and depends primarily on the problem parameters $(c,q)$ rather than specific node positions within the interval. As interval size increases beyond $c \gtrsim 5$, the curves begin to exhibit visible curvature, transitioning from linear to sub-linear growth as the slow decay of Bessel ratios at large $c$ begins to dominate convergence behavior. The complementary regime—weak $c$-dependence at high $q$—is demonstrated in Figure~\ref{fig:regime_analysis}(b), where curves at fixed high orders ($q=50, 70, 100$) show that a hundred-fold increase in interval half-width produces only mild increases in required terms, with the effect becoming progressively weaker at higher orders where the triangular recurrence structure ensures that convergence is achieved within approximately 10 terms beyond the first non-zero contribution at $n=q$.

\subsection{Ratio calculations: comparison with the McCurdy--Ng--Parlett method}
\label{sec:mccurdy_comparison}

We next compare the Chebyshev--Bessel algorithm with the standard
McCurdy--Ng--Parlett (MNP) method~\cite{McCurdy1984} for evaluating exponential
divided differences.
The comparison focuses on a capability that is essential in quantum Monte
Carlo applications but largely absent from classical approaches: the stable
and overflow-free evaluation of ratios of exponential divided
differences.

As a brief reminder, the MNP method computes divided differences using the
classical recurrence \eqref{eq:divdiff_def} starting from $f[x_i]=\exp(x_i)$.
To mitigate catastrophic cancellation when nodes are closely spaced
(\(|x_k-x_0|\lesssim0.7\)), the algorithm switches to a Taylor-series expansion.
The resulting method requires ${\cal O}(q^2\log c)$ operations and explicitly
evaluates $\exp(x_i)$ for each node.
By contrast, the Chebyshev approach first maps nodes affinely to $[-1,1]$ and
represents the exponential divided difference as a truncated
Chebyshev--Bessel expansion.
This yields a computational cost of ${\cal O}(qN)$, where the truncation length
$N$ is governed by $\max\{c,q\}$ up to the regime-dependent corrections
described in Sec.~\ref{sec:runtime_analysis}.

Both methods ultimately encounter limits imposed by floating-point overflow.
In IEEE~754 double precision, $\exp(710)\approx10^{308}$ exceeds the largest
representable value, and both algorithms fail at comparable values of $c$.
Using extended precision (\texttt{long double}, 80-bit), both methods can be
pushed to $c\sim1000$, agreeing to relative errors of order $10^{-11}$.
These limits are therefore dictated by numerical precision rather than by the
intrinsic stability of either algorithm.

The decisive advantage of the Chebyshev--Bessel formulation emerges when
computing ratios of exponential divided differences defined on the same
interval $[a,b]$ and of the same order $q$—a ubiquitous operation in quantum
Monte Carlo, where acceptance probabilities are typically expressed as ratios
of weights.

\begin{figure}[htbp]
    \centering
    \includegraphics[width=0.7\textwidth]{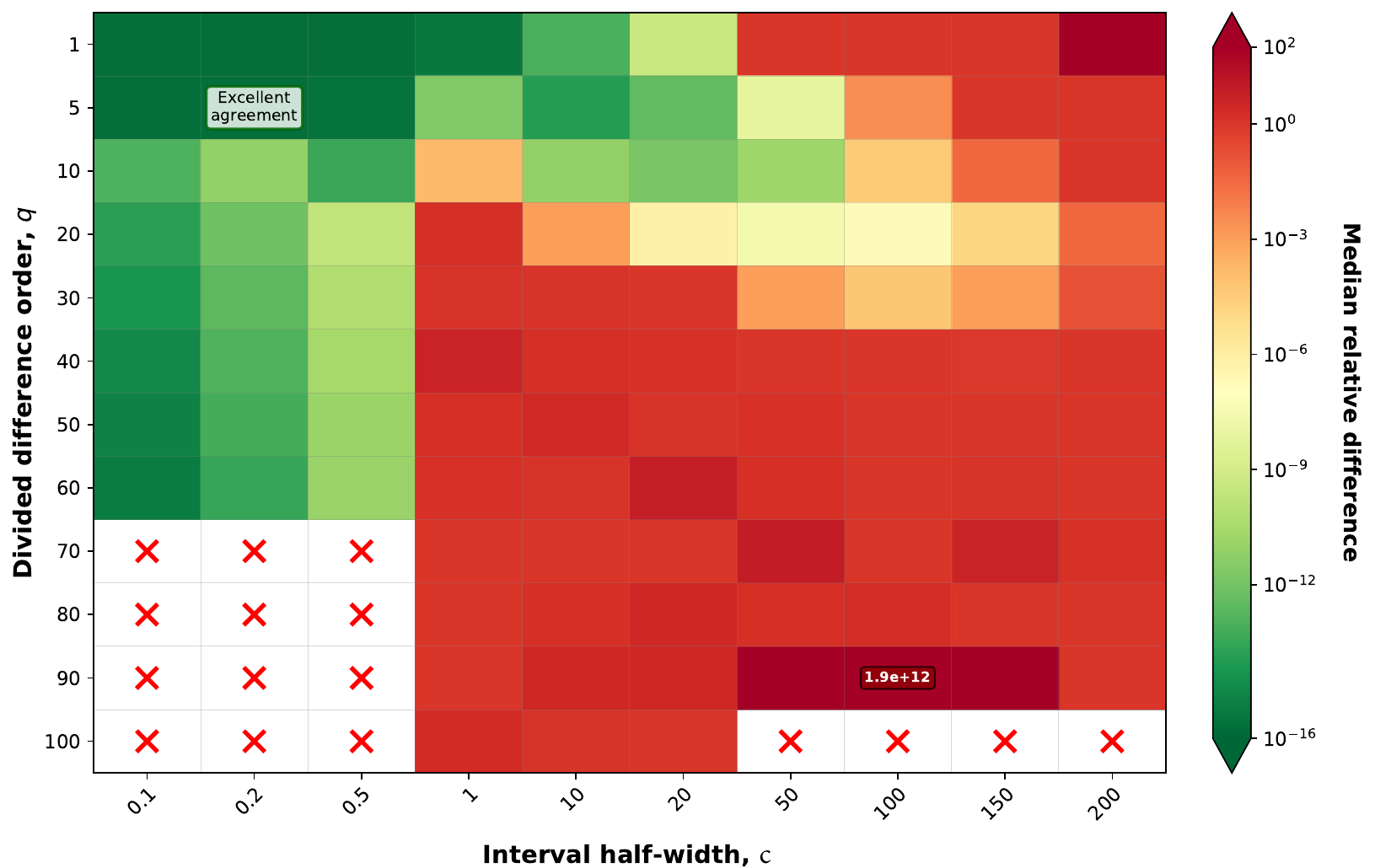}
    \caption{%
        Median relative difference between MNP and Chebyshev normalized methods
        for ratios of exponential divided differences.
        The heatmap shows
        $|\text{ratio}_{\text{Cheby}}-\text{ratio}_{\text{MNP}}|
        /|\text{ratio}_{\text{Cheby}}|$
        across $(c,q)$ parameter space, where $c$ is the interval half-width and
        $q$ is the divided-difference order.
        Colors indicate agreement on a logarithmic scale, from green
        ($<10^{-12}$) to red ($>10^{-1}$).
        Red crosses denote MNP failures, defined as cases where the algorithm
        produces a negative or non-finite result.
        Each cell represents the median over 100 random trials.
    }
    \label{fig:ratio_comparison_heatmap}
\end{figure}

For two node sets $\{x_i\}$ and $\{x'_i\}$ defined on the same interval $[a,b]$,
the ratio of exponential divided differences can be written as
\begin{equation}
R
=
\frac{\exp[x_0,\ldots,x_q]}{\exp[x'_0,\ldots,x'_q]}
=
\frac{e^{d} I_0(c)\, S_x / c^q}{e^{d} I_0(c)\, S_{x'} / c^q}
=
\frac{S_x}{S_{x'}},
\label{eq:ratio_cancellation}
\end{equation}
where $S_x$ and $S_{x'}$ are the normalized Chebyshev--Bessel sums
\[
S
=
D_q^{(0)} + 2\sum_{n=1}^{N} R_n(c)\, D_q^{(n)}.
\]
All exponential scaling factors—including $e^d$, $I_0(c)$, and $c^q$—cancel exactly.
As a result, the ratio depends only on the normalized sums $S_x$ and $S_{x'}$,
which remain ${\cal O}(1)$ to ${\cal O}(10^3)$ even when the individual
exponential divided differences exceed $10^{400}$.
No overflow is possible in the ratio evaluation, and no large intermediate quantities are ever formed.

The MNP method (or any comparable technique) admits no analogous
normalization. It must compute both exponential divided differences in
full before taking their ratio, rendering it vulnerable to overflow
whenever the individual quantities exceed representable limits—even in
cases where the ratio itself is well conditioned. This distinction is
clearly visible in Fig.~\ref{fig:ratio_comparison_heatmap}, where the
Chebyshev method remains accurate across the entire parameter space,
while MNP exhibits widespread failures at large $c$ and $q$.

A further practical advantage of the Chebyshev--Bessel formulation over
MNP and conventional table-based methods, which we mention in passing,
arises when exponential divided differences must be evaluated repeatedly
for node sets that differ by the insertion or removal of a small number
of nodes. In existing techniques, even a single-node modification
typically requires rebuilding the divided-difference table, at a cost
that scales at least linearly in $q$ and often quadratically once
numerical stabilization is taken into account. By contrast, the
Chebyshev method represents the divided difference as a truncated
expansion in Chebyshev divided differences multiplied by scalar Bessel
coefficients that depend only on the interval width. For a fixed
truncation length, these Chebyshev layers admit efficient incremental
updates when nodes are added or removed, without requiring the
reconstruction of a global divided-difference table.

\section{Incremental updates for dynamic node sets}
\label{sec:incremental}

In many applications—most notably permutation-matrix–based quantum Monte
Carlo (PMR-QMC)~\cite{EzzellBarashHen2025,BabakhaniBarashHen2025,AkaturkHen2024,AlbashWagenbrethHen2017,GuptaHen2020,GuptaAlbashHen2020,BarashBabakhaniHen2024}—
exponential divided differences must be evaluated repeatedly for node sets
that evolve dynamically through the insertion or removal of individual
nodes.
As discussed in the previous section, recomputing the full Chebyshev
divided-difference table after each such update incurs a cost of
\({\cal O}(qN)\), where \(q\) is the divided-difference order and \(N\) is
the Chebyshev truncation length.
When such updates are performed millions of times along Monte Carlo
trajectories, this cost rapidly becomes the dominant computational
bottleneck.

In this section, we show that when the affine mapping interval \([a,b]\)
is held fixed, the insertion of a single additional node admits an
incremental update whose cost is independent of \(q\).
Specifically, the updated exponential divided difference can be evaluated
in \({\cal O}(N)\) time, where \(N\) depends only on the interval
half-width \(c\) and the target accuracy.
This reduction is essential for the practical viability of PMR-based
sampling algorithms.

\subsection{Incremental Chebyshev divided-difference recurrence}

Assume that the exponential divided difference for the node set
\(x_0,\ldots,x_q\) has already been evaluated using the Chebyshev--Bessel
representation
\begin{equation}
\exp[x_0,\ldots,x_q]
=
\frac{e^{d}I_0(c)}{c^q}
\left(
D_q^{(0)} + 2\sum_{n=1}^{N} R_n(c)\,D_q^{(n)}
\right),
\label{eq:incremental_base}
\end{equation}
where \(D_q^{(n)}=T_n[y_0,\ldots,y_q]\) are the Chebyshev divided differences
of the normalized nodes.
Now suppose that a new node \(x_{q+1}\in[a,b]\) is appended, with normalized
coordinate
\[
y_{q+1}=\frac{x_{q+1}-d}{c}.
\]

The key observation is that the Chebyshev divided differences for the
augmented node set,
\[
D_{q+1}^{(n)} = T_n[y_0,\ldots,y_q,y_{q+1}],
\]
can be computed incrementally, without recomputing any previously
evaluated quantities.
Specializing the Chebyshev divided-difference recurrence
[Eq.~\eqref{eq:cheb_divdiff_rec}] to the final column yields, for all
\(n\ge2\),
\begin{equation}
D_{q+1}^{(n)}
=
2\Bigl(
y_{q+1}\,D_{q+1}^{(n-1)} + D_q^{(n-1)}
\Bigr)
-
D_{q+1}^{(n-2)}.
\label{eq:incremental_recurrence}
\end{equation}
This recurrence advances the Chebyshev order \(n\) while updating only the
new highest-order divided difference.
Each step requires (i)~the cached values \(D_q^{(n-1)}\) from the previous
node set and (ii)~the two most recent incremental values
\(D_{q+1}^{(n-1)}\) and \(D_{q+1}^{(n-2)}\).
No access to earlier divided-difference columns is required, and no
recomputation of existing values is performed.

\subsection{Incremental evaluation of the exponential divided difference}

Once the incremental Chebyshev divided differences \(D_{q+1}^{(n)}\) have
been generated, the updated exponential divided difference is obtained as
\begin{equation}
\exp[x_0,\ldots,x_q,x_{q+1}]
=
\frac{e^{d}I_0(c)}{c^{q+1}}
\left(
D_{q+1}^{(0)} + 2\sum_{n=1}^{N'} R_n(c)\,D_{q+1}^{(n)}
\right),
\label{eq:incremental_exp}
\end{equation}
where \(N'\) is the truncation index determined by the adaptive convergence
criterion.

Because each Chebyshev order \(n\) is processed in \({\cal O}(1)\) time and
the number of retained terms \(N'\) is governed by
\(\max\{c,q\}\) up to the regime-dependent corrections described in
Sec.~\ref{sec:runtime_analysis}, the total cost of the incremental update
is \({\cal O}(N')\) for fixed tolerance.
In particular, this cost is independent of the divided-difference order
\(q\).

\subsection{Incremental node removal}
\label{subsec:incremental_removal}

We now consider the inverse operation: removal of a single node from the
current node set.
Such removals arise naturally in Monte Carlo updates that delete
previously inserted energies and must be handled with efficiency
comparable to that of insertion moves.

Assume that the current node set
\(x_0,\ldots,x_q,x_{q+1}\) was obtained by appending \(x_{q+1}\) to a
previous set \(x_0,\ldots,x_q\), and that the corresponding Chebyshev
divided differences
\[
D_{q+1}^{(n)} = T_n[y_0,\ldots,y_q,y_{q+1}]
\]
were computed and cached during the incremental insertion step.
We further assume that the values
\[
D_q^{(n)} = T_n[y_0,\ldots,y_q]
\]
for the reduced node set are also available, either because they were
retained explicitly or because they can be restored from cached data
structures.

Under these assumptions, removing the node \(x_{q+1}\) requires no
recomputation of Chebyshev divided differences.
The divided differences \(D_q^{(n)}\) for the shortened node set are
already known, and the exponential divided difference is obtained
directly from the Chebyshev--Bessel expansion
\begin{equation}
\exp[x_0,\ldots,x_q]
=
\frac{e^{d}I_0(c)}{c^{q}}
\left(
D_q^{(0)} + 2\sum_{n=1}^{N''} R_n(c)\,D_q^{(n)}
\right),
\label{eq:removal_exp}
\end{equation}
where \(N''\) is the truncation index determined by the adaptive
convergence criterion for the reduced node set.

If the Chebyshev divided differences \(D_q^{(n)}\) are retained, node
removal requires only reevaluating the truncated Chebyshev sum
\eqref{eq:removal_exp}.
This costs \({\cal O}(N'')={\cal O}(c)\) operations and is independent of
the divided-difference order \(q\).
In particular, no reverse recurrence or table reconstruction is
required.

In typical PMR-QMC implementations, insertion and removal moves are
paired, and a natural strategy is to cache the Chebyshev layers
\(\{D_q^{(n)}\}_{n\le N}\) alongside each configuration.
Under this strategy, both node insertion and node removal admit
\({\cal O}(N)={\cal O}(c)\) updates, yielding a fully symmetric and
reversible incremental scheme.

Removing an interior node may always be implemented as a sequence of
last-node removals followed by reinsertions of the remaining nodes.
However, this incurs a cost proportional to the number of affected nodes
and therefore scales as \({\cal O}(qc)\) in the worst case.
The incremental algorithms presented here are thus optimized for the
last-in–first-out updates that arise naturally in reversible Monte Carlo
sampling.

\section{Evaluating scaled exponential divided differences}
\label{sec:beta_bessel}

In numerous applications, one requires the scaled exponential divided difference
\(\exp(-\beta[x_0,\ldots,x_q])\), meaning the divided difference of the function
\(f_\beta(x)=e^{-\beta x}\) evaluated at a fixed set of nodes \(x_0,\ldots,x_q\). This object arises naturally whenever exponential kernels appear with an external scale parameter, most prominently in imaginary-time quantum mechanics and finite-temperature statistical physics, where \(\beta\) plays the role of an inverse temperature. In PMR and its generalizations, for example, configuration weights are expressed directly in terms of exponential divided differences of energy sequences, with \(\beta\) controlling thermal suppression. During Monte Carlo sampling, the node set \(\{x_0,\ldots,x_q\}\) changes dynamically as off-diagonal operators are inserted or removed, while \(\beta\) remains a global control parameter. In such settings, it is neither sufficient nor convenient to work only with unscaled exponential divided differences, nor with divided differences evaluated at rescaled nodes; instead, one must evaluate \(\exp(-\beta[x_0,\ldots,x_q])\) itself, preserving the dependence on \(\beta\) at the level of the function rather than the nodes. Efficient and stable algorithms for computing scaled exponential divided differences are therefore essential for PMR-based quantum Monte Carlo methods, as well as for related applications involving thermal propagators, Euclidean-time evolution, and stiff exponential kernels.

Although the scaled exponential divided difference \(\exp(-\beta[x_0,\ldots,x_q])\) is a distinct object from the unscaled quantity \(\exp[x_0,\ldots,x_q]\), it can nevertheless be evaluated using the same underlying algorithm by exploiting the composition rule for divided differences. Specifically, if \(f(x)=\exp(x)\) and \(g(x)=f(-\beta x)=e^{-\beta x}\), then divided differences satisfy the exact identity
\[
g[x_0,\ldots,x_q] = (-\beta)^q\, f[-\beta x_0,\ldots,-\beta x_q].
\]
Thus, the divided difference of the scaled exponential at the original nodes is obtained by evaluating the unscaled exponential divided difference at the rescaled nodes \(-\beta x_0,\ldots,-\beta x_q\) and multiplying by the known prefactor \((-\beta)^q\). In practice, this means that any algorithm capable of computing \(\exp[x_0,\ldots,x_q]\)—including Chebyshev–Bessel or incremental divided-difference schemes—can be reused verbatim for the scaled kernel \(e^{-\beta x}\), with the only modifications being a preprocessing rescaling of the nodes and a final multiplicative factor. This reduction is particularly advantageous in applications such as PMR-based Monte Carlo, where \(\beta\) varies as a global parameter while the node sets evolve dynamically, allowing scaled exponential divided differences to be computed with no additional algorithmic complexity beyond that of the unscaled case.

We next discuss a key structural consequence of the Chebyshev formulation developed above as it applies to the divided difference of the scaled exponential function and show that it admits an explicit and highly economical representation as a finite linear combination of modified Bessel functions.

\subsection{Chebyshev--Bessel representation for the scaled exponential function}

\begin{theorem}[Chebyshev--Bessel representation of scaled exponential divided differences]
\label{thm:scaled_exp_cheb_bessel}
Let $f_\beta(x)=e^{-\beta x}$ with $\beta\ge 0$, and let $x_0,\ldots,x_q\in\mathbb{R}$ be a set of nodes contained in an interval $[a,b]$. Define
\[
d=\frac{a+b}{2}, \qquad c=\frac{b-a}{2},
\]
and introduce the affine map $x=d+cy$, so that the normalized nodes are
\[
y_j=\frac{x_j-d}{c}\in[-1,1], \qquad j=0,\ldots,q,
\]
whenever $c>0$. Let
\[
D_q^{(n)} := T_n[y_0,\ldots,y_q]
\]
denote the $q$th divided difference of the Chebyshev polynomial $T_n$ evaluated at the normalized nodes.

If $c>0$, the divided difference of the scaled exponential function admits the exact representation
\begin{equation}
\label{eq:bessel_beta_dd}
f_\beta[x_0,\ldots,x_q]
=
\frac{e^{-\beta d}}{c^{q}}
\left(
I_0(\beta c)\,D_q^{(0)}
+
2\sum_{n=1}^{\infty}(-1)^n I_n(\beta c)\,D_q^{(n)}
\right),
\end{equation}
where $I_n$ denotes the modified Bessel function of the first kind.

In the degenerate case $c=0$ (i.e., $x_0=\cdots=x_q=x_0$), the divided difference reduces to the confluent limit
\[
f_\beta[x_0,\ldots,x_q]
=
\frac{(-\beta)^q}{q!}\,e^{-\beta x_0}.
\]
\end{theorem}

\begin{proof}
Writing $x=d+c y$ gives $f_\beta(x)=e^{-\beta d}e^{-(\beta c)y}$. The classical
Chebyshev--Bessel expansion on $[-1,1]$,
\[
e^{t y}=I_0(t)+2\sum_{n=1}^{\infty} I_n(t)\,T_n(y),
\]
applied with $t=-\beta c$ and using $I_n(-t)=(-1)^n I_n(t)$ yields
\[
e^{-(\beta c)y}
=
I_0(\beta c)+2\sum_{n=1}^{\infty} (-1)^n I_n(\beta c)\,T_n(y).
\]
Taking divided differences on the nodes $y_0,\ldots,y_q$, invoking linearity,
and finally applying the affine scaling rule
$f[x_0,\ldots,x_q]=c^{-q}g[y_0,\ldots,y_q]$ for $g(y)=f(d+c y)$ produces
Eq.~\eqref{eq:bessel_beta_dd}.
\end{proof}

\subsection{Separation of $\beta$ dependence}

We find that for a fixed node set $x_0,\ldots,x_q$, all dependence of
$f_\beta[x_0,\ldots,x_q]$ on the rate  $\beta$ enters solely
through the scalar factors $e^{-\beta d}$ and $I_n(\beta c)$. Therefore, the coefficients
$D_q^{(n)}$ are independent of $\beta$ and may be precomputed once and reused
for all values of $\beta$. Equation~\eqref{eq:bessel_beta_dd} shows that, for fixed nodes, the function
\[
\beta \;\longmapsto\; f_\beta[x_0,\ldots,x_q]
\]
is an exponentially weighted finite linear combination of modified Bessel
functions in the single argument $\beta c$. Consequently, evaluating
$f_\beta[x_0,\ldots,x_q]$ for many $\beta$ values reduces to repeated evaluation
of $I_n(\beta c)$ with fixed coefficients. 

This separation can be contrasted with more conventional approaches for
computing exponential divided differences, which generally do not admit
an analogous decomposition into node-dependent and $\beta$-dependent
contributions. In such methods, changing $\beta$ modifies all initial
data in the divided-difference table and necessitates rebuilding the
entire structure from scratch, even when the underlying node set
$\{x_0,\ldots,x_q\}$ is held fixed. This structural coupling prevents
efficient reuse of intermediate results across multiple $\beta$ values
and renders these approaches ill suited for applications that require
evaluating $f_\beta[x_0,\ldots,x_q]$ for many rates $\beta$ on a fixed
node set.

\section{Summary and conclusions}
\label{sec:conclusions}
We have developed and analyzed a Chebyshev polynomial–based framework for the efficient and
numerically stable evaluation of exponential divided differences. By combining the
Chebyshev–Bessel expansion of the exponential function with a direct recurrence for Chebyshev
divided differences, the method avoids the catastrophic cancellation inherent to classical
divided-difference tables while exposing a transparent and predictive structure for both numerical
stability and computational complexity.

A central outcome of this work is a clear separation between the roles played by the
divided-difference order $q$ and the spectral (half-)width $c$ of the node set. We have shown that
the total computational cost factorizes into an ${\cal O}(q)$ per-term cost associated with advancing the
Chebyshev divided-difference recurrence, and a truncation length $N$ governed by the decay of
modified Bessel coefficients. At fixed numerical tolerance, this yields an overall runtime of
${\cal O}(qN)$ with $N=\Theta(c)$ in the width-dominated regime. Dependence on $q$ enters only through
the structural constraint $N\ge q$ and through subleading polynomial effects, leading to two
distinct and analytically characterizable regimes: for $q\ll c$, the truncation length is essentially
independent of $q$, while for $q\gtrsim c$ it grows linearly with $q$. This decomposition yields a
precise and implementation-independent complexity model that is borne out by extensive numerical
experiments.

Beyond single-shot evaluation, we have introduced an incremental update scheme for dynamically
evolving node sets. When the affine mapping interval is held fixed, the insertion or removal of a
single node can be handled in ${\cal O}(N)=\Theta(c)$ time, independent of the existing divided-difference
order. This eliminates the explicit ${\cal O}(q)$ factor present in full recomputation and leads to a
provable speedup proportional to $q$. The incremental formulation highlights a structural advantage
of the Chebyshev divided-difference recurrence: the Chebyshev polynomial degree and the
divided-difference order are cleanly decoupled, allowing updates to be localized to a single column
of the recurrence without reconstructing a global table.
Comprehensive numerical studies validate our analytical predictions. We verified correctness
against reference implementations at machine precision across wide parameter ranges, confirmed
the predicted scaling of the truncation length with both input set width and divided-difference order. 

The Chebyshev formulation further enables a particularly efficient treatment of scaled exponential
divided differences of the form $\exp(-\beta[x_0,\ldots,x_q])$. All dependence on the scale parameter
$\beta$ enters exclusively through scalar modified Bessel functions, while the Chebyshev divided
differences depend only on the node geometry. This separation allows intermediate results to be
reused across many values of $\beta$, a feature that is not available in classical table-based
algorithms and is especially valuable in applications involving thermal or imaginary-time
propagation.

Taken together, these results establish the Chebyshev–Bessel approach as a robust and efficient
computational kernel for exponential divided differences. The combination of analytical guarantees,
predictable scaling, numerical stability, and support for efficient incremental updates makes the
method particularly well-suited for large-scale applications in numerical linear algebra and quantum
Monte Carlo, where exponential divided differences arise repeatedly and node sets evolve dynamically.

In particular, the method directly addresses a long-standing
computational bottleneck in permutation matrix–based quantum Monte Carlo,
where exponential divided differences must be evaluated millions of
times along sampling trajectories. For single evaluations, the method
replaces the ${\cal O}(q^2 \log c)$ scaling characteristic of
table-based approaches with an ${\cal O}(q c)$ cost, reflecting a
different—and often more favorable—dependence on the divided-difference
order $q$ and the spectral half-width $c$. More importantly, for
incremental updates the cost is reduced from ${\cal O}(q^2)$, or
${\cal O}(q c)$ as in the approach presented in Ref.~\cite{GuptaAlbashHen2020}, to
${\cal O}(c)$, constituting a qualitative change in algorithmic scaling.
For typical PMR-based simulations with $q$ in the hundreds or thousands,
this improvement can translate into speedups of one to two orders of
magnitude.

Beyond quantum Monte Carlo, the algorithm provides a stable and efficient building block for any
setting requiring repeated evaluation of exponential divided differences, including matrix
exponential computations, exponential integrators for stiff differential equations, and
interpolation schemes involving clustered or dynamically changing nodes. The clean distinction
between width-controlled and order-controlled regimes offers predictable performance characteristics
that simplify algorithm design and resource estimation in large-scale scientific codes.

A full C++ reference implementation accompanying this work is publicly available~\cite{Hen2025DividedDifferenceChebyshev}.
The repository includes both the Chebyshev–Bessel algorithm and the classical McCurdy–Ng–Parlett
method, optimized routines for incremental node insertion and removal, and diagnostic tools for
accuracy and performance testing. Together, these resources provide a complete and reproducible
platform for applying, benchmarking, and extending the methods developed here.

\section*{Acknowledgments}

We thank Lev Barash for useful discussions and comments. 

\bibliographystyle{elsarticle-num}
\bibliography{refs}

\begin{thebibliography}{10}
\expandafter\ifx\csname url\endcsname\relax
  \def\url#1{\texttt{#1}}\fi
\expandafter\ifx\csname urlprefix\endcsname\relax\def\urlprefix{URL }\fi
\expandafter\ifx\csname href\endcsname\relax
  \def\href#1#2{#2} \def\path#1{#1}\fi

\bibitem{stoer2002introduction}
J.~Stoer, R.~Bulirsch, Introduction to Numerical Analysis, 3rd Edition,
  Springer-Verlag, New York, 2002.

\bibitem{hildebrand1987introduction}
F.~B. Hildebrand, Introduction to Numerical Analysis, 2nd Edition, Dover
  Publications, New York, 1987.

\bibitem{EzzellBarashHen2025}
N.~Ezzell, L.~Barash, I.~Hen, A universal black-box quantum monte carlo
  approach to quantum phase transitions, npj Computational Materials 11 (2025)
  Article 9, also available as arXiv:2408.03924.
\newblock \href {https://doi.org/10.1038/s41524-025-01891-0}
  {\path{doi:10.1038/s41524-025-01891-0}}.

\bibitem{BabakhaniBarashHen2025}
A.~Babakhani, L.~Barash, I.~Hen, A quantum monte carlo algorithm for arbitrary
  high-spin hamiltonians, arXiv preprint arXiv:2503.08039Preprint (2025).

\bibitem{AkaturkHen2024}
E.~Akaturk, I.~Hen, A quantum monte carlo algorithm for bose-hubbard models on
  arbitrary graphs, Physical Review B 109 (2024) 134519.
\newblock \href {https://doi.org/10.1103/PhysRevB.109.134519}
  {\path{doi:10.1103/PhysRevB.109.134519}}.

\bibitem{AlbashWagenbrethHen2017}
T.~Albash, G.~Wagenbreth, I.~Hen, Off-diagonal expansion quantum monte carlo,
  Physical Review E 96 (2017) 063309.
\newblock \href {https://doi.org/10.1103/PhysRevE.96.063309}
  {\path{doi:10.1103/PhysRevE.96.063309}}.

\bibitem{GuptaHen2020}
L.~Gupta, I.~Hen, Elucidating the interplay between non-stoquasticity and the
  sign problem, Advanced Quantum Technologies 3~(1) (2020) 1900108.
\newblock \href {https://doi.org/10.1002/qute.201900108}
  {\path{doi:10.1002/qute.201900108}}.

\bibitem{GuptaAlbashHen2020}
L.~Gupta, T.~Albash, I.~Hen, Permutation matrix representation quantum monte
  carlo, Journal of Statistical Mechanics: Theory and Experiment 2020 (2020)
  073105.
\newblock \href {https://doi.org/10.1088/1742-5468/ab9e64}
  {\path{doi:10.1088/1742-5468/ab9e64}}.

\bibitem{BarashBabakhaniHen2024}
L.~Barash, A.~Babakhani, I.~Hen, A quantum monte carlo algorithm for arbitrary
  spin-1/2 hamiltonians, Physical Review Research 6 (2024) 013281.
\newblock \href {https://doi.org/10.1103/PhysRevResearch.6.013281}
  {\path{doi:10.1103/PhysRevResearch.6.013281}}.

\bibitem{isaacson1994analysis}
E.~Isaacson, H.~B. Keller, Analysis of Numerical Methods, Dover Publications,
  New York, 1994.

\bibitem{McCurdy1984}
A.~McCurdy, K.~C. Ng, B.~N. Parlett, Accurate computation of divided
  differences of the exponential function, Mathematics of Computation 43~(168)
  (1984) 501--528.
\newblock \href {https://doi.org/10.1090/S0025-5718-1984-0758200-5}
  {\path{doi:10.1090/S0025-5718-1984-0758200-5}}.

\bibitem{Opitz1964}
G.~Opitz, Steigungsmatrizen, Zeitschrift f{\"u}r Angewandte Mathematik und
  Mechanik 44 (1964) 231--234.
\newblock \href {https://doi.org/10.1002/zamm.19640440526}
  {\path{doi:10.1002/zamm.19640440526}}.

\bibitem{Zivcovich2019}
F.~Zivcovich, A note on the computation of the divided differences of the
  exponential function, Dolomites Research Notes on Approximation 12 (2019)
  28--42.

\bibitem{DLMF10.41}
F.~W.~J. Olver, D.~W. Lozier, R.~F. Boisvert, C.~W. Clark, Nist digital library
  of mathematical functions, \url{https://dlmf.nist.gov/10.41}, section 10.41:
  Computation of Modified Bessel Functions (2023).

\bibitem{Hen2025DividedDifferenceChebyshev}
I.~Hen, Divided-difference exponentials with chebyshev polynomials, GitHub
  repository,
  \url{https://github.com/itay-hen/Divided-difference-exponentials-with-Chebyshev-polynomials}
  (2025).

\end{thebibliography}

\end{document}